\documentclass[11pt,letterpaper]{article}

\usepackage[margin=1in]{geometry}
\usepackage{amsmath,amssymb,amsfonts}
\usepackage{graphicx}
\usepackage{tikz}
\usetikzlibrary{positioning,arrows.meta,shapes.symbols,fit,backgrounds}
\usepackage{booktabs}
\usepackage[font=small,labelfont=bf,skip=5pt]{caption}
\usepackage{subcaption}
\usepackage{multirow}
\usepackage{longtable}
\usepackage{xcolor}
\usepackage{hyperref}
\usepackage{url}
\usepackage{listings}
\usepackage{enumitem}
\usepackage[numbers]{natbib}

\setlist[itemize]{topsep=2pt,itemsep=1pt,parsep=0pt,leftmargin=1.4em}

\lstdefinelanguage{Dafny}{
  morekeywords={method,function,lemma,returns,requires,ensures,decreases,var,
    assert,assume,if,else,while,seq,real,int,bool,then,forall,exists,
    invariant,predicate,nat,as,reads,modifies,old,ghost,datatype,match,case,
    calc,by,return,map,set,array,string,char},
  sensitive=true,
  morecomment=[l]{//},
  morecomment=[s]{/*}{*/},
  morestring=[b]",
}

\lstdefinestyle{fig}{
  frame=none,
  columns=fullflexible,
  keepspaces=true,
  xleftmargin=0pt,
  aboveskip=2pt,
  belowskip=0pt,
}

\newcommand{\sys}{NFV}
\newcommand{\daf}{\textsc{Dafny}}
\newcommand{\boog}{\textsc{Boogie}}
\newcommand{\cbmc}{\textsc{CBMC}}

\newcommand{\nag}{\textsc{Nagini}}
\newcommand{\lean}{\textsc{Lean}}
\newcommand{\fstar}{\textsc{F}$^{\star}$}

\newcommand{\anverifybad}{5}

\newcommand{\anrefutebad}{6}

\newcommand{\anverifyrawprec}{94}
\newcommand{\anverifyrawrec}{70}

\newcommand{\anrefuterawprec}{89}
\newcommand{\anrefuterawrec}{46}

\newcommand{\anbadtotal}{11}

\newcommand{\anpipepop}{206}
\newcommand{\anpipeissued}{128}

\newcommand{\anpipeprec}{92}
\newcommand{\anpiperec}{57}
\newcommand{\anpipeopen}{43}

\newcommand{\castrongpre}{3}

\newcommand{\dsonepairs}{155}

\newcommand{\dsgenmodel}{claude-sonnet-4.5}
\newcommand{\dsgentests}{100}
\newcommand{\dstested}{155}
\newcommand{\dspassed}{116}
\newcommand{\dsspecfails}{6}
\newcommand{\dsunconfirmed}{33}

\newcommand{\dstypeonly}{13}
\newcommand{\dspairs}{103}
\newcommand{\dsentries}{206}

\newcommand{\Anbadtotal}{\ifnum\anbadtotal=11 Eleven\else\anbadtotal\fi}
\newcommand{\jdgtotal}{206}

\newcommand{\jdgsaidbadpct}{76}
\newcommand{\jdgwrong}{58}

\newcommand{\jdgbuggy}{103}
\newcommand{\jdgfalsealarm}{56}
\newcommand{\jdgmissed}{2}
\newcommand{\jdgpairs}{103}

\newcommand{\jdgallprec}{72}

\newcommand{\cbmcfpall}{7}

\newcommand{\cbmccexclaims}{72}
\newcommand{\cbmccexgood}{65}
\newcommand{\cbmccexprec}{90}
\newcommand{\cbmccexrec}{63}

\newcommand{\cbmccleanall}{71}

\newcommand{\Cbmccleanall}{\ifnum\cbmccleanall=71 Seventy-one\else\cbmccleanall\fi}

\begin{document}

\title{Neuro-Formal Verification: Agentic Language-Agnostic Formal Program Reasoning}
\author{Shuvendu K.\ Lahiri\\Microsoft Research}
\date{}
\maketitle

\begin{abstract}
Formal verification provides the strongest correctness guarantees available
for software, and verification-aware languages can produce sound,
machine-checked proofs. Recent AI coding agents have dramatically lowered the
cost of constructing such proofs. Yet these benefits reach few mainstream
developers: most use languages without formal-verification support, and
formalizing properties and modeling execution environments require expertise
in formal methods. Proof therefore remains reserved for a small number of
notable artifacts that are used in practice, while production software is
attested primarily through review and testing.

We introduce \emph{neuro-formal verification} (NFV), which brings this
automation to mainstream programming languages. An AI coding agent formalizes
a source-level verification problem into a proof obligation in a
verification-aware language, which is discharged by an established sound
verifier aided by agentic proof search. NFV uses staged, goal-blind
transformations to reduce the risk of proving an artifact that does not
faithfully represent the source program, property, or environment. Since NFV
cannot ensure the soundness of the formalization --- a correspondence that can
only be loosely defined in the presence of informally stated environment
assumptions --- it optimizes for empirical accuracy rather than
end-to-end soundness, while insisting on machine-checked evidence for every
verdict.

Experiments with current frontier models on a balanced dataset of correct and
buggy solutions to Python programming problems demonstrate the effectiveness
of our approach. NFV with \daf{} correctly resolves \anpiperec{}\% of all
entries, with \anpipeprec{}\% precision among its verdicts; with a \cbmc{}
backend, it produces a counterexample for \cbmccexrec{}\% of the buggy
programs at \cbmccexprec{}\% precision. In contrast, an \emph{LLM-as-judge}
baseline achieves only \jdgallprec{}\% precision despite answering every entry
without producing any checkable artifact; an unstaged agent--verifier
combination proves 98\% of both the correct and the known-buggy programs,
yielding only 50\% precision. Together, they confirm that both proofs and
staging benefit an AI agent's formal program reasoning.
\end{abstract}

\section{Introduction}
\label{sec:intro}

Formal program verification refers to the use of mathematical logic to derive
a mathematical \emph{proof} of a formal property $\phi$ against a
formal model of a program $P$ operating in an execution environment $E$. 
The idea dates to Turing~\cite{turing1949} and
Hoare~\cite{hoare1969}, and it has been held out ever since as the strongest
available form of assurance for program correctness. Decades of work have made
it real. A range of mature verification-aware languages and proof assistants ---
\daf{}~\cite{leino2010dafny}, F$^\star$~\cite{swamy2016fstar},
Verus~\cite{lattuada2023verus}, \textsc{Lean}~\cite{demoura2021lean},
\textsc{Rocq}~\cite{rocq}, \textsc{Isabelle/HOL}~\cite{nipkow2002isabelle} ---
combine SMT-backed~\cite{barrett2016smtlib,demoura2008z3} automation with user-supplied proof
structure in varying proportions. They have been used to prove substantial
systems correct, including some that are widely used~\cite{klein2009sel4,leroy2009compcert,bhargavan2017everest,cutler2024cedar}.
Such success stories are, however, relatively few and often involve experts in
those languages and proof assistants. They are typically green-field projects
in which code and proof are co-developed under a methodology that takes years
to establish.

A proof of $\phi$ for a program $P$ in a mainstream language $L$ faces two
additional obstacles. First, $L$'s constructs must be formalized --- in a program logic or
an intermediate verification language (IVL) such as
\boog{}~\cite{barnett2006boogie}. Efforts in that direction, from
VCC~\cite{cohen2009vcc} to Verus~\cite{lattuada2023verus}, are real but slow and
cover a handful of languages. Even for a language as studied as C there is no
single verifier for the whole language, as \textsc{CBMC}~\cite{clarke2004cbmc},
VCC and Frama-C~\cite{kirchner2015framac} embody different memory models,
specification formalisms and proof systems, each catering to a different
appetite for automation. Second, a proof about $P$ is a proof about $P$ running
in an environment $E$ --- the precondition the developer has in mind, the
semantics of the libraries $P$ calls, the external state it reads and updates
--- all of which have to be inferred from documentation and world knowledge. The
\textsc{SLAM/SDV} developers, for instance, spent decades building
and maintaining accurate models of the operating system in order to verify
device drivers against it~\cite{ball2011slam}. As a result, for the code that actually
ships, developers have stuck with code review and testing as the cost-effective
way to obtain assurance.

Two developments of the past year change the calculation. First, developers
increasingly produce code with AI coding agents, a practice sometimes called
``vibe coding''~\cite{karpathy2025vibe}, which raises the demand for assurance.
Second, those same agents turn out to be unexpectedly good at writing code
\emph{and proofs} in exactly the verification-aware languages whose expert
supply has long been a
bottleneck~\cite{misu2024dafny,first2023baldur,ioannidis2026proofspromptly}.

\medskip
\noindent
This motivates the question this paper sets out to answer:
\begin{quote}
\itshape
Can we harness the proof automation of mature verification-aware languages
for mainstream developers in a language-agnostic way?
\end{quote}

\textbf{Our central conjecture is that} a technique capable of distinguishing
\emph{correct programs} (those that satisfy a stated property in a given
environment) from \emph{incorrect programs} (those that do not), with extremely
high empirical accuracy --- say, 99.99\% --- while
producing a checkable artifact that explains each verdict could be invaluable
for auditing code, understanding it, and calibrating trust in it, especially
when the code is AI-generated.

In this paper, we describe \emph{neuro-formal verification} (NFV) as one
possible way to realize this vision. An effective neuro-formal verifier would
allow a developer using a mainstream programming language $L$ to pose
\emph{formal reasoning questions} (say, as formal assertions in $L$) and
obtain answers that are (a)~empirically accurate without necessarily providing
end-to-end guarantees of soundness or completeness,
(b)~produced at the push of a button by default, and (c)~accompanied by a proof
over formal models of the program, the property, and the environment.

\subsection{Neuro-formal verification: the whole picture}
\label{sec:northstar}

\paragraph{The inputs.} A developer has a program $P_L$ in some source language
$L$; a property $S_{nl}$ they care about, written in whatever notation they
already use --- a docstring, an issue, a comment; and an environment $E$ in
which the program executes --- including the context that calls it and the
libraries and services it calls --- described by documentation $E_{nl}$.

\begin{figure}[t]
\centering
\includegraphics[width=\linewidth]{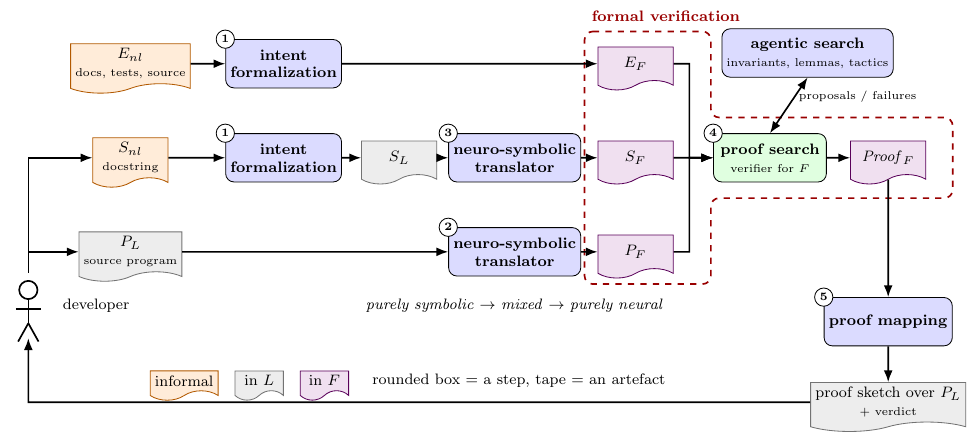}
\caption{Neuro-formal verification. Circled numbers are the five stages
enumerated above. Rounded boxes transform; tape-shaped nodes are the
artifacts they consume and produce, colored by the language the artifact lives
in --- orange for informal English, gray for the source language $L$, purple for
the formal language $F$. Blue is a step an LLM performs, green a step a verifier
performs. The developer authors $S_{nl}$ and $P_L$ and reads the proof
sketch that comes back. The dashed red region is the classical formal
verification problem --- given $P_F$, $S_F$ and $E_F$, find a proof.}
\label{fig:northstar}
\end{figure}

\paragraph{The workflow.} Figure~\ref{fig:northstar} lays out five stages that
turn those inputs into a checked proof over a model of a program and return it
in the developer's own terms.
The workflow is parameterized by a verification-aware language $F$ with
first-class formal verification support.

\begin{enumerate}
\item \textbf{Intent formalization}~\cite{lahiri2026intent}.
  Synthesize from $E_{nl}$ a formal specification $E_F$ in $F$ --- the
  contracts of the libraries the program calls and any preconditions stated
  by the documentation --- and from $S_{nl}$ an assertion $S_L$ in the source
  language that expresses what the developer meant. Both are claims about intent, and
  both should be \emph{attested} against every artifact available --- the
  documentation, the tests, and the source when available.
\item \textbf{Program translation}, $P_L \rightarrow P_F$, into the formal
  language $F$ of some mature verifier. This step can 
  range from purely symbolic (a compiler, where a faithful front end exists)
  through mixed approaches to purely neural (this paper).
\item \textbf{Specification translation}, $S_L \rightarrow S_F$, along the same
  spectrum.
\item \textbf{Proof search over $F$}, discharged by the verifier for $F$ and
  driven agentically: an agent proposes invariants, lemmas, and tactics; the
  verifier checks each proposal, so the search may use neural methods freely
  without putting the verifier's soundness at risk.
\item \textbf{Proof mapping.} Map a successful formal proof
  $\mathit{Proof}(P_F, S_F, E_F)$ into a \emph{proof sketch} stated over $P_L$,
  using the developer's own variables and naming the assumptions in terms of
  $E_{nl}$.
\end{enumerate}

The workflow should not be construed as end-to-end formal verification that
$P_L$ satisfies $S_L$. The
source-language semantics, the environment description, and the translations
are not all formally attested. We therefore use the term \emph{formal program
reasoning}.

\paragraph{Proof beyond the final verdict.} Formal proof is not confined to
establishing $S_F$ for $P_F$ under $E_F$. Intent formalization can also produce
a test suite $T_F$, either extracted from library artifacts or generated by
agents, together with proofs of the soundness, completeness, and preciseness of
$E_F$ with respect to $T_F$~\cite{lahiri2026intent}. Similarly, the path from
$S_{nl}$ to $S_F$ can produce tests and proofs attesting $S_F$. These artifacts
strengthen the attestation of the formalized environment and specification,
even though they do not establish end-to-end correctness.
More generally, the explicit formal artifacts $P_F$, $S_F$, and $E_F$ give
users inspectable accounts of the program model, the property being checked,
and the assumptions under which the result holds. A user can review these
artifacts, identify mismodeling, and provide feedback that improves subsequent
formalization.
Users can also vet familiar concrete tests rather than inspect $S_F$ and $E_F$
directly, reducing the formal-methods expertise required of them.

\subsection{An example}
\label{sec:example}

Let us illustrate the workflow for this paper with a simple two-line Python program.
We restrict the list elements to be exact rationals
(Python's \texttt{fractions.Fraction}), so that the division operator \texttt{/} performs no rounding ---
an assumption needed for the specification to hold.

\begin{center}
\begin{minipage}{0.74\linewidth}
\begin{lstlisting}[language=Python,basicstyle=\ttfamily\footnotesize,columns=fullflexible,keepspaces=true]
def shifted_mean(l):
    "Average the list after adding 1 to each element; l must be non-empty."
    return sum(x + 1 for x in l) / len(l)

# postcondition
assert shifted_mean(l) == sum(l) / len(l) + 1
\end{lstlisting}
\end{minipage}
\end{center}

\noindent
Informally, the correctness of this specification relies on
 \[
  \mathit{len}(l) \neq 0
 \] 
and 
\[
  \mathit{sum}(\,\{x+1 \mid x \in l\}\,) \;=\; \mathit{sum}(l) + \mathit{len}(l).
\]

Stage~5 would not only return ``verified'' but also the proof-sketch reasoning, in natural language: \emph{the specification
holds because adding $1$ to each of \texttt{len(l)} elements adds
\texttt{len(l)} to their sum, provided the list is non-empty} --- an artifact the developer can check.

\paragraph{The same example, worked in \daf{}.} Instantiating $F$ with \daf{}
makes the workflow concrete. Stage~1 starts from $E_{nl}$, which for
this program is the documentation of Python's \texttt{sum}\footnote{The
built-in's runtime docstring --- \texttt{sum.\_\_doc\_\_} in CPython~3.14 --- quoted
as it stands, less a closing sentence restricting it to numeric values.} --- all that is
actually on offer about the library:

\begin{center}
\begin{minipage}{\linewidth}
\begin{lstlisting}[basicstyle=\ttfamily\normalsize,columns=fullflexible,keepspaces=true,frame=none]
Return the sum of a 'start' value (default: 0) plus an iterable of numbers.
When the iterable is empty, return the start value.
\end{lstlisting}
\end{minipage}
\end{center}

\noindent
Stage~1 formalizes this documentation as $E_F$, which in this example consists
only of a recursive definition of \texttt{Sum}:

\begin{center}
\begin{minipage}{0.86\linewidth}
\begin{lstlisting}[language=Dafny,basicstyle=\ttfamily\footnotesize,columns=fullflexible,keepspaces=true]
function Sum(l: seq<real>): real {
  if |l| == 0 then 0.0 else l[0] + Sum(l[1..]) }
\end{lstlisting}
\end{minipage}
\end{center}

\noindent
\texttt{Sum} is attestable against the docstring: it is a recursive definition
whose base and recursive cases can be read directly against the documented
behavior.

Stages~2 and~3 then render $P_F$ and $S_F$. Translating the generator expression
\texttt{x + 1 for x in l} introduces \texttt{Shift} as part of the model of
$P_F$; it is not part of $E_F$:

\begin{center}
\begin{minipage}{0.86\linewidth}
\begin{lstlisting}[language=Dafny,basicstyle=\ttfamily\footnotesize,columns=fullflexible,keepspaces=true]
function Shift(l: seq<real>): seq<real> {
  seq(|l|, i requires 0 <= i < |l| => l[i] + 1.0) }

method ShiftedMean(l: seq<real>) returns (r: real)
  requires |l| > 0
  ensures r == Sum(l) / (|l| as real) + 1.0
{ ShiftedSum(l);
  var n := |l| as real;
  r := Sum(Shift(l)) / n;
  DivDistrib(Sum(l), n); }
\end{lstlisting}
\end{minipage}
\end{center}

\noindent
The two calls in the method are proof steps, not part of $P_F$. Stage~4
introduces both lemmas. \texttt{ShiftedSum} proves by induction that translating
the generator preserves the expected relationship between the two sums.
\texttt{DivDistrib} supplies the nonlinear arithmetic step that the verifier
does not close automatically when $n$ is symbolic:

\begin{center}
\begin{minipage}{0.86\linewidth}
\begin{lstlisting}[language=Dafny,basicstyle=\ttfamily\footnotesize,columns=fullflexible,keepspaces=true]
lemma ShiftedSum(l: seq<real>)
  ensures Sum(Shift(l)) == Sum(l) + |l| as real
{ if |l| > 0 {
    assert Shift(l)[1..] == Shift(l[1..]);
    ShiftedSum(l[1..]); } }

lemma DivDistrib(a: real, n: real)
  requires n != 0.0
  ensures (a + n) / n == a / n + 1.0
{ assert n / n == 1.0;
  assert (a + n) / n == a / n + n / n; }
\end{lstlisting}
\end{minipage}
\end{center}

\noindent
\daf{} checks both lemmas and, with their calls in
\texttt{ShiftedMean}, discharges all seven obligations of the resulting file.
Stage~5 would map this proof back to the proof-sketch reasoning stated above:
\emph{the specification holds because adding $1$ to each of \texttt{len(l)}
elements adds \texttt{len(l)} to their sum, provided the list is non-empty}.
The current instantiation evaluated in this paper does not include this mapping.

\paragraph{The same example with a bug, refuted.} 
Suppose the
developer had written the reduction as a single addition on the total rather
than one per element --- \texttt{r := (Sum(l) + 1.0) / n} in place of
\texttt{r := Sum(Shift(l)) / n}, a one-token slip.

\begin{center}
\begin{minipage}{0.74\linewidth}
\begin{lstlisting}[language=Python,basicstyle=\ttfamily\footnotesize,columns=fullflexible,keepspaces=true]
def shifted_mean(l):
    "Average the list after adding 1 to each element; l must be non-empty."
    return (sum(l) + 1) / len(l)  # bug
\end{lstlisting}
\end{minipage}
\end{center}

\noindent
Now $P_F$ does not meet
$S_F$, and stage~5 should say so. The tempting move is to report the verifier's
failure: \daf{} cannot prove the postcondition.
However, a proof failure cannot be interpreted as a bug --- an unproved goal is evidence about
the proof search, not about the program.

What makes a refutation attestable is the same thing that makes a verification
attestable --- a proof that $S_F$ fails. The program is left untouched, $S_F$ is
negated, and the witness is added as a precondition, so the verifier is asked to
confirm a positive claim rather than to fail at a negative one:

\begin{center}
\begin{minipage}{0.86\linewidth}
\begin{lstlisting}[language=Dafny,basicstyle=\ttfamily\footnotesize,columns=fullflexible,keepspaces=true]
method ShiftedMean(l: seq<real>) returns (r: real)
  requires |l| > 0
  requires l == [0.0, 0.0]                      // the witness
  ensures !(r == Sum(l) / (|l| as real) + 1.0)  // S_F, negated
{ var n := |l| as real;
  r := (Sum(l) + 1.0) / n;                      // the bug
  assert Sum(l) == l[0] + Sum(l[1..]);
  assert Sum(l[1..]) == l[1] + Sum(l[1..][1..]);
  assert l[1..][1..] == []; }
\end{lstlisting}
\end{minipage}
\end{center}

\noindent
The witness $l = [0,0]$ makes the buggy program return $1/2$ where
$S_F$ demands $1$; the three asserts are the proof-repair loop unfolding
\texttt{Sum} on that input, and \daf{} discharges the negated goal. 
On the Python list \texttt{[Fraction(0), Fraction(0)]} the buggy program itself returns
\texttt{Fraction(1, 2)} where the specification demands \texttt{Fraction(1)}, and the
\texttt{assert} raises \texttt{AssertionError}.

\subsection{Contributions of this paper}
\label{sec:whatwedo}
This paper provides the first empirical instantiation of the NFV framework
described in Figure~\ref{fig:northstar}. Specifically, we contribute:

\begin{itemize}
\item A dataset of \dsentries{} Python (program, specification) pairs --- \dspairs{}
correct references and \dspairs{} buggy mutants of them --- derived from prior
work~\cite{endres2024nl2postcond}.
\item The \sys{} framework, including its goal-blind staged discipline and the
gates and checks that enforce it.
\item Two instances of the framework: a proof-based pipeline using \daf{} and a
bounded-verification pipeline using \cbmc{}.
\item An empirical comparison against an \emph{LLM-as-judge} baseline and an
unstaged \emph{agent--verifier} baseline, isolating the contributions of proof and
of staging.
\end{itemize}

\paragraph{Scope of this instantiation.} We begin at $S_L$, because the
benchmark supplies an assertion synthesized from a docstring by prior
work~\cite{endres2024nl2postcond}. We synthesize $E_F$ but do \emph{not}
formally attest it against tests describing $E_{nl}$, and both translation
stages sit at the neural end of the spectrum. Finally, we stop at a verdict
backed by a machine-checked proof in $F$ rather than map that proof into a
sketch over $P_L$; evaluating such sketches requires a careful user study.

\noindent
Within this scope, we demonstrate encouraging and timely results:

\begin{itemize}
\item With a flagship frontier model, 
\begin{itemize}
  \item the \daf{} backend proves \textbf{\anverifyrawrec{}\%} of the correct programs and refutes
\textbf{\anrefuterawrec{}\%} of the
mutants at \anverifyrawprec{}\% and \anrefuterawprec{}\% precision, signifying the viability of the approach.
  \item the combined pipeline issues a machine-checked verdict on
\anpipeissued{} of the \anpipepop{} entries, correctly resolving
\textbf{\anpiperec{}\%} of all entries at \textbf{\anpipeprec{}\%} precision.
\item \cbmc{} refutes \textbf{\cbmccexrec{}\%} of the mutants at
\textbf{\cbmccexprec{}\%} precision --- the highest recall and precision for
finding bugs in this dataset.
 \item On this dataset, a majority of incorrect verdicts stem from incorrect precondition synthesis in $E_F$ that can be improved with
better intent-formalization methods.
\end{itemize}
\item Less capable frontier models perform significantly worse on both verification and refutation --- this capability 
may not have existed even a year ago.
\item NFV outperforms the strawman \emph{LLM-as-judge} baseline on this dataset: the judge answers
every entry at \jdgallprec{}\% precision
(compared to the \anpipeprec{}\% precision NFV provides through its ability to abstain when it cannot
back its verdict with a proof artifact). 
\item NFV also outperforms an unstaged \emph{agent--verifier} baseline, which proves
98\% of the correct programs \emph{and} 98\% of the buggy ones at 50\% precision,
confirming the benefit of the staged discipline.
\end{itemize}

\paragraph{A dual reading.} We have described the architecture from the
perspective of a program verifier: a formal method that uses an LLM as its front end. It reads
equally well from the other side, as \emph{equipping an AI agent with a formal
verifier}: the LLM is the agent, the pipeline is a tool it calls to discharge
a claim about program correctness. 

\paragraph{Organization.} Section~\ref{sec:dataset} describes the dataset,
Section~\ref{sec:nfv} the framework, Sections~\ref{sec:eval}
and~\ref{sec:analysis} the experiments and the errors that remain,
Section~\ref{sec:related} the related work, and Section~\ref{sec:concl}
concludes with directions for future work.\footnote{The dataset, the pipeline
and the scripts that produce every table and number reported here will be
released shortly as open source.}

\section{Dataset}
\label{sec:dataset}

This section describes a dataset we curate to study neuro-formal verification
empirically.

\subsection{Curation}

\textbf{Source.} We select the programs and specifications from the
\textsc{nl2postcond} benchmark~\cite{endres2024nl2postcond}.\footnote{\url{https://github.com/microsoft/intent-formalization/tree/main/nl2postcondition-fse2024}}
This dataset builds on HumanEval~\cite{chen2021humaneval}, whose tasks provide
(i) a natural-language \emph{docstring} stating the intended
behavior of the function, together with a handful of worked input/output
examples, and (ii) a \emph{reference}
Python implementation $P$, taken to be correct with respect to a set of tests $T$
augmented by \textsc{EvalPlus}~\cite{liu2023evalplus}.
It augments each task with (iii) a set of postconditions synthesized from the task docstring and
(iv) a set of mutants $M$ that fail the tests.
Each postcondition is labeled \emph{sound} if it passes each test in $T$ over
$P$; its \emph{completeness} is the fraction of $M$ killed by running the test
inputs in $T$ with the postcondition asserted.
We take the postconditions from that repository's
\path{eval_results/gpt4_simple_no_ref.zip} --- the GPT-4 \textsc{SimplePromptNoRef}
arm, whose \path{summary_eval_results.jsonl} carries the soundness and
completeness verdicts and whose
\path{individual_postcondition_results/} directory carries one file per
candidate --- and the mutants together with their killing inputs from
\path{eval_results/distinct_code_mutants.jsonl.zip}.

\textbf{Initial pool.}
For each task, we choose the specification that has the highest completeness score.
Next, we choose the first mutant that the specification is confirmed to kill on
an input in $T$.
This provides us with \dsonepairs{} reference implementations (labeled correct)
and \dsonepairs{} mutant implementations (labeled incorrect), where each pair
shares a specification $S$.

\textbf{Filtering.} We then apply two filters. The first checks
whether the reference fails the specification --- the goal is to 
perform additional testing to
remove any example where the specification is incorrect, assuming 
the reference implementation is correct. 
\dsgenmodel{} was prompted for \dsgentests{} diverse inputs per
task and these were executed against $P$ and $S$ over all \dstested{}
references. We dropped \dsspecfails{} tasks where $P$ was observed to violate
$S$, and \dsunconfirmed{} more whose tests did not run to completion, leaving
\dspassed{}. Our approach is conservative ---
a falsifying input may itself violate a precondition the docstring states only
in prose.
The second filter drops
\dstypeonly{} tasks whose postcondition constrains only the return \emph{type},
an obligation that a typed target language $F$ encodes in the method signature and
that therefore does not exercise the program verifier. 
What remains is \textbf{\dsentries{}
entries} over \dspairs{} tasks, balanced by construction.
Appendix~\ref{sec:specs} lists every one of them with the postcondition it is
scored against and the upstream identifiers that pin it, so the corpus can be
reconstructed without rerunning the filtering steps that are not
deterministic.

\begin{table}[t]\centering\small
\caption{\texttt{nl2post-safe} by library tier.}
\label{tab:dataset}
\begin{tabular}{llrrr}
\toprule
Tier & What the program touches & Tasks & Entries & Share \\
\midrule
L0 & nothing to axiomatize & 21 & 42 & 20\% \\
L1 & total built-ins only & 22 & 44 & 21\% \\
L2a & collections, first-order only & 25 & 50 & 24\% \\
L2b & a function passed as an argument & 26 & 52 & 25\% \\
L3 & module imports, unmodeled built-ins & 9 & 18 & 9\% \\
\midrule
\textbf{All} & & \textbf{103} & \textbf{206} & \textbf{100\%} \\
\bottomrule
\end{tabular}
\end{table}

\subsection{Characteristics}
Table~\ref{tab:dataset} and
Table~\ref{tab:size}(\subref{tab:size-loc})--(\subref{tab:size-control})
characterize the resulting corpus along two axes: the library surface each
program uses, and its size and control flow.

\textbf{Tiers.} Table~\ref{tab:dataset} splits the corpus by \textbf{library
tier} L0--L3, a function of the calls the Python source makes: a proxy for how
much the LLM must formalize the environment, not for algorithmic
difficulty. \textbf{L0} needs no axiom at all --- arithmetic, indexing, control
flow, \texttt{len}, \texttt{range}; \textbf{L1} adds total built-ins
(\texttt{abs}, \texttt{sum}, \texttt{ord}); \textbf{L2a} is first-order
collection use; \textbf{L2b} passes a function as an argument, so the
specification must quantify over a body it cannot see; \textbf{L3} is a module
import or any call outside these. Appendix~\ref{sec:tierex} shows an example of
each. 

\textbf{Size.} Table~\ref{tab:size}(\subref{tab:size-loc}) gives program size in
terms of lines of code (LOC). The programs are small, in part because Python's
rich built-in and library functionality allows substantial behavior to be
expressed in few lines. Table~\ref{tab:size}(\subref{tab:size-control})
summarizes control flow: a majority contain a loop and most a branch.
The mutants are independently written solutions
rather than minimal edits and are often longer than their references. 

\begin{table}[t]\centering\small
\caption{Size and control flow of the 206 programs. LOC counts distinct physical lines of the
function under test carrying a syntax node; comments, blank lines,
imports and the docstring are excluded. Paired by task, mutants are
longer than their reference on 73 of 103 tasks and
shorter on 18, a median gap of 2 lines.}
\label{tab:size}
\begin{subtable}{\linewidth}
\centering
\begin{tabular}{lrrrrrrr}
\toprule
 & min & p25 & median & mean & p75 & p90 & max \\
\midrule
LOC & 2 & 4 & 7 & 7.6 & 10 & 12 & 29 \\
\bottomrule
\end{tabular}
\caption{Program size.}
\label{tab:size-loc}
\end{subtable}

\smallskip
\begin{subtable}{\linewidth}
\centering
\begin{tabular}{lrr@{\qquad}lrr}
\toprule
control flow & \# & \% & control flow & \# & \% \\
\midrule
loop-free & 92 & 45\% & $\ge 1$ branch & 144 & 70\% \\
$\ge 1$ loop & 114 & 55\% & $\ge 1$ comprehension & 37 & 18\% \\
$\ge 2$ loops & 24 & 12\% & & &  \\
\bottomrule
\end{tabular}
\caption{Control flow.}
\label{tab:size-control}
\end{subtable}
\end{table}

\section{Neuro-formal verification (NFV)}
\label{sec:nfv}

This section instantiates the neuro-formal verification framework (Figure~\ref{fig:northstar}) 
by elaborating some of the components. 
First, we describe some syntactic \emph{safeguards} to ensure that the AI agent does not simplify or alter the 
verification task midway through the pipeline (e.g., changing the translated source $P_F$ during neural proof search); these are collected in Section~\ref{sec:verifproof}.
Second, we describe how to adapt the framework to perform a proof of refutation of the specification $S_F$ (Section~\ref{sec:refutation}).
Finally, we also extend the framework to accommodate bounded-program verifiers (such as \textsc{CBMC} or \textsc{Corral}) that search 
for witnesses over bounded unrolling of the program loops and recursion (Section~\ref{sec:cbmcarm}). 

\subsection{The proof-based pipeline}
\label{sec:proofbased}

We outline both the verification and refutation arms of this pipeline. 
Both directions roughly run the same sequence of steps and hand the result to a deductive
verifier; they differ only in the goal. The steps refine the stages of
Figure~\ref{fig:northstar}: the library theory (stage~1a) and the precondition (stage~1b) are
the two halves of $E_F$ produced by stage~1; transliterating the body is
stage~2; rendering the specification is stage~3; and proof search and repair are
stage~4. 
We instantiate the framework with a specific choice of $L$ (Python) and $F$ (\daf{}), but the concepts should 
be broadly applicable to other programming languages (C\#, Java, JavaScript) and verifier backends (\fstar{}, \lean{}).
We choose \daf{} as our target verification-aware language for this work since it supports many 
features of modern imperative programming languages such as sequences and so lets the 
translation stay close to the source.

\subsubsection{Verification proof}
\label{sec:verifproof}

The verification arm aims to produce a proof of $S_L$
over a translated program. 

Stage~1a formalizes the Python library surface the program touches and stage~1b
synthesizes preconditions from the docstring; both freeze before the
specification is visible, so neither the library theory nor the precondition can
be chosen to suit the goal.

\paragraph{Provenance tags.} Stage~2 tags every emitted line with either
\texttt{src:}, quoting its Python line verbatim, or \texttt{synth:}, marking it
as having no source counterpart (declarations, invariants, \texttt{decreases}
clauses). Untagged code is rejected --- without that rule, rewritten logic hides
in lines the audit cannot see. Translation has its own repair loop, but the only
feedback is to run a syntax checker (e.g., \texttt{dafny resolve} in \daf{}): 
type and resolution errors. The loop sees no verifier goal, so it has no information with which to
make the program \emph{correct}, only enough to make it well-formed.

Figure~\ref{fig:example} shows the partial artifact for our running example.
Every line carries a provenance tag, and the two \texttt{src:} lines
quote the Python verbatim. \texttt{Sum} is the only stage-1a content here:
what Python's \texttt{sum} means, written before the goal is visible.
\emph{Frozen} marks
what proof search may not change --- every \texttt{src:} line, the stage-1a
theory, the stage-1b precondition and the goal itself --- each re-checked
mechanically on every repair round; the lemmas and the two proof calls at the
bottom are what the repair loop contributed.
\texttt{Shift} encodes the generator expression and so belongs to the
translation that calls it. 
\texttt{ShiftedSum} and \texttt{DivDistrib}
are the output of proof search --- the facts the goal reduces to, found from the
verifier's failures --- and both are lemmas \emph{proved} in the target rather
than assumed. The full artifact verifies at seven obligations.

\begin{figure}[t]
\centering
\begin{minipage}[t]{0.30\textwidth}
\footnotesize\textbf{Python + specification}
\begin{lstlisting}[language=Python,style=fig]
def shifted_mean(l):
    "Average the list after
     adding 1 to each
     element; l must be
     non-empty."
    return sum(x + 1
      for x in l) / len(l)

assert return_value ==
       sum(l)/len(l) + 1
\end{lstlisting}
\end{minipage}\hfill
\begin{minipage}[t]{0.66\textwidth}
\footnotesize\textbf{Translated \daf{} artifact}
\begin{lstlisting}[language=Dafny,style=fig]
// synth: frozen library theory
// doc: sum(iterable, start=0) returns start plus the
// items of the iterable; on an empty iterable, start.
function Sum(l: seq<real>): real
{ if |l| == 0 then 0.0 else l[0] + Sum(l[1..]) }

// synth: translation helper, for the generator expression
function Shift(l: seq<real>): seq<real>
{ seq(|l|, i requires 0 <= i < |l| => l[i] + 1.0) }

// synth: proof, proved not assumed
lemma ShiftedSum(l: seq<real>)
  ensures Sum(Shift(l)) == Sum(l) + |l| as real
{ /* induction on l */ }
lemma DivDistrib(a: real, n: real) requires n != 0.0
  ensures (a + n) / n == a / n + 1.0
{ /* two asserts, both checked */ }

// src: source.py:1 | def shifted_mean(l):
method shifted_mean(l: seq<real>) returns (return_value: real)
  // synth: frozen precondition, from "non-empty"
  requires |l| > 0
  // synth: frozen postcondition
  ensures return_value == Sum(l) / (|l| as real) + 1.0
{
  // src: source.py:3 | return sum(x + 1 for x in l) / len(l)
  return_value := Sum(Shift(l)) / (|l| as real);
  // synth: proof
  ShiftedSum(l); DivDistrib(Sum(l), |l| as real);
}
\end{lstlisting}
\end{minipage}
\caption{The running example of Section~\ref{sec:example} as the pipeline
emits it. The bodies of \texttt{ShiftedSum} and \texttt{DivDistrib} are elided
for space; both are proved, and Section~\ref{sec:example} gives those proofs.}
\label{fig:example}
\end{figure}

In addition to the above checks to ensure the translated code and environmental specifications
are not altered to achieve a proof, we also ensure that the precondition is not inconsistent and 
the translation does not introduce unproven assumptions. 

\subsubsection{Refutation proof}
\label{sec:refutation}

Refutation asks the opposite question --- is there an input class on which the
specification \emph{provably} fails --- and reuses stages 1a, 1b and 2
byte-identically. The goal is negated
\emph{mechanically}, the postcondition $S_F$ becomes $\neg S_F$.
Next, the LLM proposes a violating input looking at only the
Python source; that guess is installed as an additional precondition (this is an exception 
to the verification arm, where preconditions cannot be strengthened); 
and proof repair proves the negated goal with the body still frozen. 
Similar to the verification proof, this may require the synthesis of invariants, lemmas 
and intermediate assertions. 

We also check a few additional facts specific to refutation.
\textsc{Reachability} requires that replacing the postcondition with 
\emph{false} fails verification. \textsc{Goal shape} requires that only the inversion of 
$S_F$ is present as a postcondition. 

\subsection{The bounded-verification-based pipeline}
\label{sec:cbmcarm}

The second instantiation drops the proof and searches for the witness directly.
It shares most of the stages, except that they address only refutation. 
A bounded model checker (such as \textsc{CBMC}, which operates over C) 
answers the refutation question
natively: it searches for an execution violating the goal assertion within a
fixed unwinding, so a violating trace is a refutation and \emph{every} other
outcome, including a clean run, is an abstention. This arm also has no
counterpart to agentic proof search or proof repair.

\subsection{Proof-based versus bounded verification for refutation}
\label{sec:incomplete}

Both pipelines can refute a program: the proof-based one by proving that a
witnessing precondition forces a violation, the bounded one by finding a trace
that exhibits it. Neither is complete, and the challenges each faces are
different.\footnote{We assume throughout that we cannot simply run $P$ instead.
Executing $P$ may be unavailable or not cost-effective --- $P$ may be a
component of a larger system (say a device driver) needing hardware, a build or
a fixture the auditing tool does not have --- or $\phi$ may speak of
\emph{ghost state} with no runtime representation. Combining \sys{} with testing
opens up other interesting optimization possibilities not covered in this paper.}

\emph{Proof scales; search enumerates.} 
A refutation (proof) search can scale to large modules due to its modular nature when paired with an
effective (agentic) proof search automation. 
On the other hand, bounded verification is inherently non-scalable and is known to deteriorate with the depth of call chains.
Ideas such as stratified inlining~\cite{lal2012corral}, that keeps callees abstract
until a candidate trace demands them, may offer some relief --- but cannot help with indefinite scaling.

\emph{Proof is easier to inspect.} A proof is written in the vocabulary of the
translated program, so the provenance tags map it back to the source line by
line (Figure~\ref{fig:example}). A bounded counterexample is a trace, and in our
experience an interprocedural trace spanning several tens of methods is hard to 
act on unless a concrete test can be generated from it.

\emph{Proof is incomplete under internal non-determinism.} The witness is
installed as a precondition, so the obligation is that \emph{every} execution it
admits violates the specification. A bug that manifests on some choice of internal non-determinism 
--- a hash order, a clock, a service reply --- cannot be
discharged. Approaches based on \emph{Incorrectness logic}~\cite{ohearn2020incorrectness}
can alleviate this issue and also avoid unrolling loops with underapproximate loop summaries. 

\emph{Bounded search failures can be spurious.} For bounded verifiers, it is difficult to 
distinguish true violations from the ones admitted due to the environment $E_F$ being under-constrained.
On the other hand, a proof of refutation (whenever present) is a guarantee of a bug under the 
stated environment assumptions. 

\section{Evaluation}
\label{sec:eval}

\subsection{Setup}
\label{sec:knobs}

In this section we evaluate the efficacy of the NFV pipeline on our dataset. 
Unless stated otherwise the front end is \textsc{gpt-5.6-sol}, OpenAI's flagship
model at the time of writing, and the target is \daf{}.
Two budget parameters govern the proof search: an \emph{attempt} discards the current
translation and restarts from stage~1, while a \emph{repair round} attempts to repair
the proof failure through stage~4 only. 
All results use \textbf{5 attempts} and \textbf{10 repair rounds}, unless otherwise stated.

We measure \emph{precision} as the number of correct verdicts over verdicts
issued, and \emph{recall} as the number of correct verdicts issued over the entries of that class.
Abstention is not counted as a verdict.
$F1$ is their harmonic mean, which we report because the two are traded against
each other: abstaining more raises precision and lowers recall.

\begin{table}[t]\centering\small
\caption{\daf{}, the reference backend, on each task and its own population. Verification and refutation run at 5 attempts $\times$ 10 proof rounds. The pipeline row is composed from the two arms above --- refutation first, verification on what it leaves open --- so every row is the same backend. ``wrong'' counts verdicts contradicting the label.}
\label{tab:tasks}
\begin{tabular}{lrrrrrr}
\toprule
task & population & issued & wrong & prec.\ (\%) & rec.\ (\%) & $F_1$\\
\midrule
verification ($P \models S$) & 103 & 77 & 5 & 93.5 & 69.9 & 0.80\\
refutation ($P \not\models S$) & 103 & 53 & 6 & 88.7 & 45.6 & 0.60\\
\midrule
pipeline (refute, then verify) & 206 & 128 & 10 & 92.2 & 57.3 & 0.71\\
\bottomrule
\end{tabular}
\end{table}

\begin{table}[t]\centering\footnotesize
\caption{The two tasks by library tier, one run each.}
\label{tab:tiers-final}
\begin{tabular}{lrrrrrr}
\toprule
& \multicolumn{3}{c}{verification} & \multicolumn{3}{c}{refutation} \\
\cmidrule(lr){2-4}\cmidrule(lr){5-7}
Tier & proved & \emph{bad} & \emph{prec} &
refuted & \emph{bad} & \emph{prec} \\
\midrule
L0 & 14/21 (67\%) & 0 & 100\% & 10/21 (48\%) & 1 & 91\% \\
L1 & 18/22 (82\%) & 2 & 90\% & 11/22 (50\%) & 0 & 100\% \\
L2a & 15/25 (60\%) & 1 & 94\% & 12/25 (48\%) & 2 & 86\% \\
L2b & 22/26 (85\%) & 0 & 100\% & 13/26 (50\%) & 1 & 93\% \\
L3 & 3/9 (33\%) & 2 & 60\% & 1/9 (11\%) & 2 & 33\% \\
\midrule
\textbf{All} & \textbf{72/103} (70\%) & 5 & \textbf{94\%} & \textbf{47/103} (46\%) & 6 & \textbf{89\%} \\
\bottomrule
\end{tabular}
\end{table}

\subsection{The proof-based arm: \daf{}}
\label{sec:evalverify}

Table~\ref{tab:tasks} gives the result on each task using the \daf{} verifier backend.
Of the 206 (program, specification) pairs, comprising the correct references and the incorrect mutants, the 
verification arm (row marked ``verification'') generates a correctness proof in \daf{} for 72 of the 103 correct reference implementations and issues proofs for 5 of the mutants labeled incorrect.
This corresponds to a $\approx$\textbf{70\%} recall and $\approx$\textbf{93.5\%} precision.
On the other hand, the refutation workflow (row marked ``refutation'') produces a proof of violation for 47 of the 103 mutants while proving 6 reference implementations incorrect on some input.
This corresponds to a $\approx$\textbf{46\%} recall and $\approx$\textbf{89\%} precision.
In addition, we create another configuration (``pipeline'') where we first run the refutation, and then run verification if refutation does not produce a proof.
The combined pipeline produces a proof (of refutation or of verification) for 128 of the 206 tasks, marking a recall of \textbf{57.3\%} (118) and precision of \textbf{92.2\%} (118 / 128), 
and with an overall F1 score of \textbf{0.71}.

Table~\ref{tab:tiers-final} breaks the same run down by library tier.
A few points worth noting:
\begin{itemize}
  \item NFV struggles most with examples in the L3 tier for both the verification and refutation tasks, both in terms of precision and recall.
  The two have different causes.
  Recall is low because the imported libraries are hard to model, so the proof search more often gives up.
  The wrong verdicts are not caused by that modeling: Section~\ref{sec:analysis} traces all four to preconditions the 
  docstring never stated explicitly and to the argument types the translation chose.
  \item On the other hand, NFV does well in both L0 and L2b tiers, each with a single incorrect claim.
  L0 benefits from the reduced need to model libraries; 
  L2b benefits from the fact that higher-order functions such as \texttt{filter} can be easily mapped to \daf{}'s comprehension support.
  \item Finally, within L2a, programs that directly manipulate strings prove at
  4/10, compared with 11/15 for the remaining programs, suggesting that
  reasoning about Python string semantics in \daf{} remains challenging.
\end{itemize}

\subsubsection{Analysis}
\label{sec:analysis}

\Anbadtotal{} claims contradict a ground-truth label: \anverifybad{} proofs of
buggy programs and \anrefutebad{} counterexamples against correct ones. 
We investigated each of these and Table~\ref{tab:causes} attributes each to a cause.

\subsubsection{Incorrect refutations}
We assume that every reference implementation is correct with respect
to the intent of its docstring, as we did when constructing the dataset in
Section~\ref{sec:dataset}.
A counterexample against a reference therefore leaves two possibilities: either the
specification does not soundly formalize the docstring, or NFV mismodels the source
code or its environment.
Distinguishing the two can be subjective, because the intent of some docstrings is
ambiguous.
For example, the docstring for entry \texttt{49} (\texttt{modp}) documents 
``Return $2^n$ modulo $p$ (be aware of numerics)'', which does not say whether $p$ must be positive 
or $n$ non-negative.
We are conservative: we blame NFV whenever there is \emph{any} interpretation of the
docstring under which the specification is sound.

% generated by nfv/paper/check_causes.py -- do not edit by hand
\begin{table}[t]\centering\footnotesize
\caption{Every wrong verdict of the reported run, by cause. \emph{entry} is the kind of program the claim was made about: a \emph{reference} is labeled correct and was refuted, a \emph{mutant} is labeled incorrect and was proved. \emph{unstated assumption} is the precondition the specification would have needed for the verdict and the label to agree. Every row is executed rather than asserted.}
\label{tab:causes}
\begin{tabular}{llll}
\toprule
task & entry & cause & unstated assumption \\
\midrule
\texttt{HumanEval\_11} & reference & missing precondition & $|a| = |b|$ \\
\texttt{HumanEval\_20} & reference & missing precondition & no \texttt{NaN} \\
\texttt{HumanEval\_49} & reference & missing precondition & $p > 0$ \\
\texttt{HumanEval\_64} & reference & missing precondition & \textsc{ascii} \\
\texttt{HumanEval\_79} & reference & missing precondition & $\mathit{decimal} \ge 0$ \\
\texttt{HumanEval\_111} & reference & missing precondition & tokens are single letters \\
\texttt{HumanEval\_116} & mutant & missing precondition & $arr[i] \ge 0$ \\
\texttt{HumanEval\_6} & mutant & stronger precondition & no empty space-delimited field \\
\texttt{HumanEval\_133} & mutant & stronger precondition & \texttt{lst} is integer-typed \\
\texttt{HumanEval\_151} & mutant & stronger precondition & \texttt{lst} is integer-typed \\
\texttt{HumanEval\_35} & mutant & library formalization & --- \\
\bottomrule
\end{tabular}
\end{table}

We arrive at the following conclusion: all \anrefutebad{} of them
are real Python counterexamples to the postcondition as written, on inputs the
docstring never excluded. 
For each, we name the missing precondition (underlined) and explain briefly how
it blocks the buggy input.

\begin{itemize}
\item \texttt{11} (\texttt{string\_xor}): \underline{$|a| = |b|$}. The postcondition asserts
  $|\mathit{rv}| = |a| = |b|$, so on operands of unequal length it is
  unsatisfiable by \emph{any} implementation.
\item \texttt{20} (\texttt{find\_closest\_elements}): \underline{No \texttt{NaN}}. The
  witness is a pair of equal elements with $\neg(x \le x)$ --- admitted because
  \texttt{PyFloat} is abstract, and realizable because every comparison with
  \texttt{NaN} is false.
\item \texttt{49} (\texttt{modp}): \underline{$p > 0$}. The witness is a negative modulus
  ($n=0$, $p=-1$), which the theory's \texttt{requires b != 0} admits and Python
  computes without complaint.
\item \texttt{64} (\texttt{vowels\_count}): \underline{\textsc{ascii}-only \texttt{s}}.
  Python lowercases the non-\textsc{ascii} Latin capital I with dot above,
  İ (\texttt{"\textbackslash u0130"}), to \texttt{i} followed by a combining dot,
  so the specification counts one vowel, whereas the implementation tests the
  original character against an \textsc{ascii} vowel set and counts none.
\item \texttt{79} (\texttt{decimal\_to\_binary}):
  \underline{\texttt{decimal} is nonnegative}. For \texttt{-3}, Python returns \texttt{"-0b11"};
  slicing from index 2 leaves \texttt{"b11"}, so the function returns
  \texttt{"dbb11db"}, which does not match
  \texttt{\textasciicircum db[01]+db\$}.
\item \texttt{111} (\texttt{histogram}): \underline{Tokens are single letters}. The docstring
  ``Given a string representing a space separated lowercase letters'' seems to imply
  that a violating input such as \texttt{"a aa"} should not be considered. 
\end{itemize}

\subsubsection{Incorrect verifications}

\paragraph{Spurious proofs come from a stronger precondition.} 
\castrongpre{} of the \anverifybad{} are valid
theorems about a program more constrained than the Python one.

\begin{itemize}
\item \texttt{6} (\texttt{parse\_nested\_parens}) has a \emph{stronger precondition}. 
The docstring, in full, is ``Input to this function is a
  string represented multiple groups for nested parentheses separated by
  spaces....'' The inferred precondition includes a clause --- \emph{the string is nonempty, and no space is leading, trailing
  or doubled} --- that rules out the buggy inputs \texttt{''}, \texttt{'()~'},
  \texttt{'~()'} and \texttt{'()~~()'} on which the mutant fails.
\item \texttt{133} (\texttt{sum\_squares}) and \texttt{151} (\texttt{double\_the\_difference}): a \emph{stronger precondition} manifested as an incorrect \emph{type} signature:
  the parameter \texttt{lst} in Python is translated as \texttt{seq<int>} in \daf{}, which rejects the floating-point input 
  on which the mutant fails.
\end{itemize}

\paragraph{Incorrect library formalism:} \label{sec:unsound}\texttt{35} (\texttt{max\_element}) 
computes the maximum of a list, and the specification asserts
\texttt{return\_value == max(l)}, where \texttt{max} is the Python built-in.
The mutant adds 1 to the return value, violating that specification. 
The translated obligation is \texttt{result == PyMax(l)}, and \texttt{PyMax} is a
function stage~1 wrote by transcribing the mutant's own loop, its base case adding 
1 to the result and so introducing the same bug into the specification. 

\paragraph{Incorrect label:}
Finally, \texttt{116} (\texttt{sort\_array}) is an entry where NFV changes the \emph{label} of the mutant.
The docstring for this example says ``an array of non-negative
integers'' and the mutant fails the specification only for negative
values. 
NFV installs a correct precondition and is therefore able to prove that the mutant satisfies the postcondition. 
The problem came not from the \textsc{nl2postcond} dataset, but from the \textsc{EvalPlus} dataset, which added a test case containing a negative integer that violates the precondition. 

\subsection{The bounded-verifier arm: \textsc{CBMC}}
\label{sec:evalcbmc}

In this section, we discuss instantiating NFV with \cbmc{}, a bounded program verifier, 
taking C as the formal language. 
Because \cbmc{} does not support rich specification constructs, the agent writes 
C code for the libraries from their documentation. 
Given the bounded nature of the verification, we only focus on the effectiveness of 
this arm for refutation. 
Besides compiling without errors, the translated C program must be free of
\emph{undefined behavior (UB)}, which \cbmc{}'s built-in UB checks establish:
overflow, bounds, pointer and conversion checks.

\cbmc{} issues \cbmccexclaims{} counterexamples, of which \cbmccexgood{} refute a
mutant program and \cbmcfpall{} are false: that is a recall of
\textbf{\cbmccexrec{}\%} over the \dspairs{} buggy programs at
\textbf{\cbmccexprec{}\%} precision, against the \daf{} arm's
\anrefuterawrec{}\% at \anrefuterawprec{}\% on the same programs. 
This offers an alternative for refutation when bounded verification can scale. 
The precondition is again the dominant
source of unsound refutations, the same diagnosis Section~\ref{sec:analysis}
reaches for the proof arm using \daf{}. 

\subsection{Weaker LLMs}
\label{sec:weakmodel}

Both arms were re-run with three further LLMs in place of
\texttt{gpt-5.6-sol}, holding prompts, checks, backend and budget fixed.

\begin{table}[t]\centering\footnotesize
\caption{Comparison of different LLMs for the proof-based tasks. \emph{proved} is over the 103 correct programs, \emph{refuted} over the 103 buggy ones, \emph{bad} counts claims contradicting the label, \emph{prec} is precision over claims, and \emph{F1} combines that precision with recall over the 103, counting an abstention as a wrong answer would count. \emph{type} and \emph{audit} split the entries that never reached the proof step: code that failed the type checker, and code that resolved but failed the artifact checks. Rows are ordered by decreasing verification \emph{F1}. A row marked $^{\dagger}$ was stopped before finishing: its refutation half is scored over the 203 entries that completed, not 206. The 3 that did not finish are the ones the model spent longest on.}
\label{tab:weakmodel}
\begin{tabular}{lrrrrrrrrrr}
\toprule
& \multicolumn{4}{c}{verification} & \multicolumn{4}{c}{refutation} & \multicolumn{2}{c}{no proof attempted} \\
\cmidrule(lr){2-5}\cmidrule(lr){6-9}\cmidrule(lr){10-11}
& proved & \emph{bad} & \emph{prec} & \emph{F1} & refuted & \emph{bad} & \emph{prec} & \emph{F1} & \emph{type} & \emph{audit} \\
\midrule
\texttt{gpt-5.6-sol} & 72/103 (70\%) & 5 & 94\% & 0.80 & 47/103 (46\%) & 6 & 89\% & 0.60 & 1 & 9 \\
\texttt{gpt-5-mini} & 56/103 (54\%) & 38 & 60\% & 0.57 & 34/103 (33\%) & 17 & 67\% & 0.44 & 72 & 20 \\
\texttt{claude-sonnet-4.6}$^{\dagger}$ & 41/103 (40\%) & 12 & 77\% & 0.53 & 39/101 (39\%) & 2 & 95\% & 0.55 & 0 & 159 \\
\texttt{claude-haiku-4.5} & 11/103 (11\%) & 2 & 85\% & 0.19 & 7/103 (7\%) & 0 & 100\% & 0.13 & 92 & 257 \\
\bottomrule
\end{tabular}

\end{table}

We observe that:
\begin{itemize}
  \item \texttt{gpt-5.6-sol} is the only model that can provide value for this
  dataset, with the combined pipeline achieving \textbf{\anpipeprec{}\%}
  precision and \textbf{\anpiperec{}\%} recall.
  \item \texttt{claude-haiku-4.5} performs quite poorly on both tasks.
  It also has the largest number of examples where the translated \daf{} code failed to type check or pass the audits.
  \item There is a sharp drop in the verification capability of weaker models compared to \texttt{gpt-5.6-sol} 
due to either reduced recall (\texttt{claude-sonnet-4.6}) or reduced precision (\texttt{gpt-5-mini}) or both.
Both of these models have substantially more typechecking and audit failures.
  \item Degradation is task-asymmetric, and affects the verification tasks more than the refutation tasks.

\end{itemize}

\subsection{Strawman and symbolic baselines}
\label{sec:naive}

We also compare NFV against three baselines spanning a spectrum from fully neural to fully symbolic.

\emph{LLM-as-judge.} In this baseline, the LLM
is shown the program and the specification in Python --- with the docstring's input
assumptions supplied --- and asked whether the program
satisfies the specification. It commits on every one of the \jdgtotal{} entries,
never abstaining, and refutes \jdgsaidbadpct{}\% of a corpus. Overall, it is wrong 
on \jdgwrong{} of the \jdgtotal{} entries at \jdgallprec{}\% precision
(compared to the \anpipeprec{}\% precision NFV provides through its ability to abstain when it cannot
back its verdict with a proof artifact):
it rejects \jdgfalsealarm{} of the \jdgpairs{} correct programs and accepts \jdgmissed{} of the
\jdgbuggy{} incorrect ones. 
The failure is not only the error rate but what is left behind: an opinion, with no checkable artifact.

\emph{Unstaged agent--verifier.} The second baseline asks for a
verifiable \daf{} method carrying the specification, with a repair loop against
the verifier with no restrictions --- one call, no separated stages, nothing written
blind, no freeze and no checks. It proves 98\% of the correct programs \emph{and}
98\% of the known-wrong ones: precision 50\%. However, the program and specification being proved 
are not the intended ones,
since the agent is incentivized merely to find a proof: the code is rewritten
(57\% of unsound proofs), the goal is restated using invented library helpers
(35\%), or the proof rests on an explicit assumption (8\%). These categories
are mutually exclusive; additionally, 56\% of the proofs contain an unjustified
precondition, overlapping the primary categories. A proof is only as good
as the artifact it is about; so NFV provides the discipline that fixes the artifact, not the
proof search.

\emph{A sound verifier.} We also explored \nag{}, a
symbolic Python verifier~\cite{eilers2018nagini} that leverages SMT solvers for proofs. Given a
type-annotated program with contracts, it discharges obligations comparable to
the ones \daf{} discharges here. In its default configuration it proves \textbf{0\%} on this corpus, because \textbf{83\%} of the 
entries are rejected for the lack of type-annotated signatures, which these programs do not carry.
Asking an agent to supply good guesses at these annotations raises the fraction that reach the verifier to
98\%, but raises the proof rate only to \textbf{2\%}. 
This result is not a judgment of the capability of \nag{} on this dataset, but rather a demonstration of the utility of NFV.
We attribute the low success rate to the absence of
a formal environment specification $E_F$ (including preconditions that are vital even for tier L0 without any libraries) 
that symbolic verifiers assume, and to the absence of proof
automation --- precisely the burden that symbolic verification tools place on their end users.

\section{Related work}
\label{sec:related}
Neuro-formal verification is inspired by decades of investment in symbolic
program verification, its persistent challenges, and recent progress in AI agents.
We describe a few (far from exhaustive) strands of prior work in these two domains.

\subsection{Symbolic program verification}
\textbf{Translation-based verifiers:} Verifying a program by translating it
into an (intermediate) verification language is a long-standing practice.
\textsc{Spec\#}~\cite{barnett2004specsharp} translates an
annotated dialect of C\#, \textsc{VCC}~\cite{cohen2009vcc} and \textsc{Havoc}~\cite{lahiri2008havoc}
translate C, \textsc{VeriSol}~\cite{wang2019verisol} translates
Solidity, and
\daf{}~\cite{leino2010dafny} compiles its own source language, all of them into
\boog{}~\cite{barnett2006boogie} and from there into SMT. 
\textsc{CSLib}~\cite{barrett2026cslib} is a proposal for developing Lean libraries for the semantics of
languages and program logics, which will make these translations more trusted and reusable.

\textbf{Extended-static checkers:} 
NFV is inspired by works on \emph{extended static checking}~\cite{flanagan2002escjava}: the
verification is performed on a formal model of the source code, even though the formalism may be 
optimistic or ``soundy''~\cite{livshits2015soundiness}.
Such a translation either targets a fragment of the language or adopts
optimistic assumptions that trade soundness for coverage of the code it can
analyze.
Tools (such as SDV, HAVOC, etc.) of this genre have been deployed on production software for
``proofs'' of properties in systems code against a formal model of code~\cite{vanegue2013audit,ball-vstte12,ball2011slam,lal2012corral},
where the value delivered lies in hard-to-find defects that an engineer confirms and fixes.

\textbf{Environment modeling:} 
Most formal verification efforts work on closed programs and face the problem of modeling the 
environment of an \emph{open program}, which leaves end-to-end soundness either in the hands of the
user~\cite{priya2024envassume} or in heuristic inference of environment assumptions like angelic 
verification~\cite{das2015angelic,lahiri2020angelic}.
Angelic verification and the earlier work on almost-correct
specifications~\cite{blackshear2013acs} were similarly motivated by the goal of
a proof-based empirical separator of correct and incorrect programs, obtained
purely through abductive inference~\cite{dillig2012abduction} of environment
assumptions.

We envision NFV helping to unlock the value of formal verification for mainstream developers without being
constrained to specific languages and domains by leveraging progress in AI agents along three directions
discussed below. 

\subsection{Agentic program reasoning}
 
\textbf{Neural language translation} --- the problem of translating across
languages predates the current generation of
agents and LLMs~\cite{roziere2020transcoder,roziere2022unittests,szafraniec2023compiler}
but has advanced quickly with them~\cite{yang2024unitrans}.
NFV will benefit from defining datasets for translating from mainstream languages 
into verification-aware languages, where we ensure equivalence through tests or 
symbolic means. 

\textbf{Intent formalization} --- the process of generating and validating 
formal program specifications from informal documentation such as docstrings~\cite{lahiri2026intent}, 
a concept related to \emph{autoformalization}~\cite{wu2022autoformalization} from formal mathematics.
These specifications can be attested through symbolic validation against
a program's tests~\cite{lahiri2024symbolic} without the need for implementations,
to ensure their soundness, completeness and preciseness (modulo tests).
Such work was previously performed manually, but has benefited from agentic proof automation~\cite{swamy2026spot}.
This will alleviate the need for end-users of NFV to inspect non-trivial formal library specifications 
at the cost of inspecting concrete tests in the source language that are much more familiar to them. 

\textbf{Agentic proof automation} --- progress has been rapid in invariant synthesis with 
language models~\cite{kamath2024loopy,wu2024lemur}, whole-proof generation and
repair~\cite{first2023baldur}, retrieval-augmented tactic
prediction~\cite{yang2023leandojo}, and proof generation for
\textsc{Verus}~\cite{yang2025autoverus} and \daf{}~\cite{misu2024dafny}. 
We foresee continued benefits from improvements in frontier-model capability.

\section{Conclusion and Future Work}
\label{sec:concl}

We presented neuro-formal verification as a principled integration of AI coding agents 
and verification-aware languages to facilitate language-agnostic formal program reasoning.
We envision NFV not only as a cost-effective way to find defects, but also as
a means for developers to build confidence in their code gradually and to
understand the assumptions required for it to behave correctly.
We have demonstrated that recent progress in agentic reasoning is beginning to
enable such formal program reasoning:
the pipeline correctly resolves \anpiperec{}\% of a dataset at
\anpipeprec{}\% precision, backing every verdict it issues with a
machine-checked proof.
The dataset remains challenging for the current flagship model, which leaves the remaining
\anpipeopen{}\% unresolved.

Much research needs to be done to bring this to mainstream software developers not versed in formal 
verification.
We need to curate \emph{labeled} datasets that are increasingly representative of real-world 
software development tasks covering richer libraries and concurrency.
Increasing the trustworthiness of each agentic component through symbolic techniques
will be helpful.
For example, a benchmark of code translation into 
verification-aware languages, evaluated through tests or other symbolic means, can help make this 
component robust.
Similarly, we need real-world benchmarks for intent formalization, and techniques to
evaluate the quality of agent-generated specifications for the environment against
existing artifacts such as tests~\cite{lahiri2026intent,lahiri2024symbolic}.
These advances should be grounded in realistic user studies to understand the utility of the approach,
including the evaluation of informal concepts such as proof sketches.
Finally, NFV needs to be integrated with modern software-development workflows.
Lean Squad's work~\cite{syme2026leansquad} on generating Lean
proofs from source code (without necessarily adopting NFV's staged approach to
isolating sources of unsoundness) can be viewed as an experiment in this space.

NFV does not replace proof with empirical evaluation: every verdict from the
proof-based pipeline is backed by a machine-checked proof over formal models.
Empirical evaluation instead measures how faithfully those models represent
the source program, property, and environment. This end-to-end accuracy can be
compared across foundation models and improved using data-driven methods, as
for any learned component. This framing may offer a more holistic basis for
bringing formal reasoning tools to mainstream developers.

\section*{Acknowledgments}

The author is grateful to Thomas Ball for detailed feedback on an earlier
preprint of this paper.

\appendix

\section{Dataset details}
\label{sec:dsdetails}

\subsection{The benchmark in full}
\label{sec:specs}

Table~\ref{tab:specs} lists every task in \texttt{nl2post-safe} with the
postcondition its two entries are scored against, together with the upstream
identifiers that pin the pair.

\begin{footnotesize}
\begin{longtable}{@{}r l r r p{0.52\linewidth}@{}}
\caption{The 103 pairs --- all 206 entries --- with the
postcondition each pair is scored against, and the upstream identifiers that pin
it. \textbf{task} is the HumanEval number. \textbf{p} is the
\texttt{response\_num} selecting the postcondition in
\texttt{summary\_eval\_results.jsonl} and
\texttt{raw\_preprocessed\_samples.jsonl}. \textbf{m} is the
\texttt{response\_num} selecting the mutant in
\texttt{distinct\_code\_mutants.jsonl}; below 200 it is a natural mutant, at
200 and above an artificial one. The reference and the mutant of a task share a
postcondition, so one row covers two entries. Postconditions are reproduced
after sanitization (Appendix~\ref{sec:dsdetails}), which is the form the
pipeline sees.}
\label{tab:specs} \\
\toprule
\textbf{task} & \textbf{entry point} & \textbf{p} & \textbf{m} & \textbf{postcondition} \\
\midrule
\endfirsthead
\toprule
\textbf{task} & \textbf{entry point} & \textbf{p} & \textbf{m} & \textbf{postcondition} \\
\midrule
\endhead
\bottomrule
\endfoot
1 & \texttt{separate\_paren\_groups} & 2 & 0 & \texttt{assert \allowbreak{}all(\allowbreak{}(\allowbreak{}s.\allowbreak{}count(\allowbreak{}'\allowbreak{}(\allowbreak{}'\allowbreak{})\allowbreak{} \allowbreak{}=\allowbreak{}=\allowbreak{} \allowbreak{}s.\allowbreak{}count(\allowbreak{}'\allowbreak{})\allowbreak{}'\allowbreak{})\allowbreak{} \allowbreak{}and \allowbreak{}'\allowbreak{}(\allowbreak{})\allowbreak{}'\allowbreak{} \allowbreak{}in \allowbreak{}s \allowbreak{}for \allowbreak{}s \allowbreak{}in \allowbreak{}return\_value)\allowbreak{})\allowbreak{}} \\
2 & \texttt{truncate\_number} & 0 & 202 & \texttt{assert \allowbreak{}0 \allowbreak{}<\allowbreak{}=\allowbreak{} \allowbreak{}return\_value \allowbreak{}<\allowbreak{} \allowbreak{}1} \\
3 & \texttt{below\_zero} & 0 & 200 & \texttt{assert \allowbreak{}return\_value \allowbreak{}=\allowbreak{}=\allowbreak{} \allowbreak{}True \allowbreak{}and \allowbreak{}any(\allowbreak{}(\allowbreak{}sum(\allowbreak{}operations[\allowbreak{}:\allowbreak{}i \allowbreak{}+\allowbreak{} \allowbreak{}1]\allowbreak{})\allowbreak{} \allowbreak{}<\allowbreak{} \allowbreak{}0 \allowbreak{}for \allowbreak{}i \allowbreak{}in \allowbreak{}range(\allowbreak{}len(\allowbreak{}operations)\allowbreak{})\allowbreak{})\allowbreak{})\allowbreak{} \allowbreak{}or \allowbreak{}return\_value \allowbreak{}=\allowbreak{}=\allowbreak{} \allowbreak{}False} \\
4 & \texttt{mean\_absolute\_deviation} & 1 & 213 & \texttt{assert \allowbreak{}0 \allowbreak{}<\allowbreak{}=\allowbreak{} \allowbreak{}return\_value} \\
\end{longtable}
\end{footnotesize}

\section{An example program from each library tier}
\label{sec:tierex}

\paragraph{L0 --- HumanEval/157.} It uses nothing beyond arithmetic, indexing and control flow. Nothing outside the program has to be described, so the model's trusted base can be empty.
\begin{lstlisting}[language=Python]
def right_angle_triangle(a, b, c):
    return a ** 2 + b ** 2 == c ** 2 or a ** 2 + c ** 2 == b ** 2 or b ** 2 + c ** 2 == a ** 2
\end{lstlisting}

\paragraph{L1 --- HumanEval/121.} Beyond arithmetic, indexing and control flow it uses \texttt{sum}. Each call here is one self-contained, total function whose meaning the model can state in a line.
\begin{lstlisting}[language=Python]
def solution(lst):
    return sum((lst[i] for i in range(len(lst)) if i % 2 == 0 and lst[i] % 2 == 1))
\end{lstlisting}

\paragraph{L2a --- HumanEval/120.} Beyond arithmetic, indexing and control flow it uses \texttt{.sort}, \texttt{sorted}. A faithful specification has to relate whole collections --- what the result contains, and in what order --- rather than give a single value.
\begin{lstlisting}[language=Python]
def maximum(arr, k):
    return sorted(sorted(arr)[::-1][:k])
\end{lstlisting}

\paragraph{L2b --- HumanEval/30.} Beyond arithmetic, indexing and control flow it uses \texttt{filter}, \texttt{list}. The specification must quantify over an argument whose body it cannot see, which is a different demand from describing a collection.
\begin{lstlisting}[language=Python]
def get_positive(l: list):
    return list(filter(lambda x: x > 0, l))
\end{lstlisting}

\paragraph{L3 --- HumanEval/133.} Beyond arithmetic, indexing and control flow it uses \texttt{import math}, \texttt{map}, \texttt{math.ceil}, \texttt{sum}. The boundary is not the model's to choose: it must axiomatise a surface defined elsewhere, and any axiom it writes about that surface is assumed, not checked.
\begin{lstlisting}[language=Python]
def sum_squares(lst):
    import math
    return sum(map(lambda x: math.ceil(x) ** 2, lst))
\end{lstlisting}

\end{document}